\documentclass[11pt]{article}
\usepackage{amssymb}
\usepackage{amsmath}
\usepackage{amscd}
\usepackage{latexsym}

\catcode`@=11
\renewcommand{\section}{\@startsection{section}{1}{0pt}{\medskipamount}
{\medskipamount}{\large\bf}}
\numberwithin{equation}{section}
\catcode`@=12
\def\a{\alpha}
\def\b{\beta}

\def\de{\delta}
\def\eps{\epsilon}
\def\ve{\varepsilon}
\def\vk{\varkappa}

\def\vp{\varphi}
\def\vs{\varsigma}
\def\th{\theta}

\newcommand{\C}{\mathbb C}
\newcommand{\R}{\mathbb R}

\newcommand{\Gcal}{{\cal G}}
\newcommand{\Acal}{{\cal A}}

\newcommand{\Ical}{{\cal I}}
\newcommand{\Mcal}{{\cal M}}
\newcommand{\Fcal}{{\cal F}}

\newcommand{\Ncal}{{\cal N}}

\newcommand{\Pcal}{{\cal P}}

\newcommand{\gfrak}{{\mathfrak g}}
\newcommand{\mfrak}{{\mathfrak m}}
\newcommand{\hfrak}{{\mathfrak h}}

\newcommand{\ah}{{\hat{\smash{a}}}}
\newcommand{\bh}{{\hat{\smash{b}}}}

\newcommand{\tR}{{\tilde{R}}}

\def\tr{\textrm{tr}}
\def\diff{\textrm{d}}
\def\pa{\mbox{$\partial$}}
\def\sfrac#1#2{{\textstyle\frac{#1}{#2}}}
\def\+{\dagger}
\def\={\ =\ }
\def\und{\qquad\textrm{and}\qquad}
\def\and{\quad\textrm{and}\quad}
\def\with{\quad\textrm{with}\quad}
\def\for{\quad\textrm{for}\quad}

\begin{document}

\begin{titlepage}
\setcounter{page}{0}

\hspace{2.0cm}

\begin{center}

{\huge\bf
Dual infrared limits of 6d $\cal N$=(2,0) theory }

\vspace{12mm}

{\LARGE
Olaf Lechtenfeld${}^{\+\times}$ \ and \  Alexander D. Popov${}^\+$
}\\[10mm]

\noindent ${}^\+${\em
Institut f\"ur Theoretische Physik,
Leibniz Universit\"at Hannover \\
Appelstra\ss{}e 2, 30167 Hannover, Germany
}\\
{Email: alexander.popov@itp.uni-hannover.de}
\\[5mm]
\noindent ${}^\times${\em
Riemann Center for Geometry and Physics,
Leibniz Universit\"at Hannover \\
Appelstra\ss{}e 2, 30167 Hannover, Germany
}\\
{Email: olaf.lechtenfeld@itp.uni-hannover.de}

\vspace{20mm}

\begin{abstract}
\noindent
Compactifying type $A_{N-1}$ 6d $\Ncal{=}(2,0)$ supersymmetric CFT on a product manifold 
$M^4\times\Sigma^2=M^3\times\tilde{S}^1\times S^1\times\Ical$ either over $S^1$ or over $\tilde{S}^1$ 
leads to maximally supersymmetric 5d gauge theories on $M^4\times\Ical$ or on $M^3\times\Sigma^2$, respectively. 
Choosing the radii of $S^1$ and $\tilde{S}^1$ inversely proportional to each other, these 5d gauge theories are dual 
to one another since their coupling constants $e^2$ and $\tilde{e}^2$ are proportional to those radii respectively.
We consider their non-Abelian but non-supersymmetric extensions, i.e.~SU($N$) Yang--Mills theories on $M^4\times\Ical$ 
and on $M^3\times\Sigma^2$, where $M^4\supset M^3=\R_t\times T_p^2$ with time~$t$ and a punctured 2-torus, 
and $\Ical\subset\Sigma^2$ is an interval. In the first case, shrinking $\Ical$ to a point reduces to Yang--Mills theory 
or to the Skyrme model on~$M^4$, depending on the method chosen for the low-energy reduction.  In the second case, 
scaling down the metric on $M^3$ and employing the adiabatic method, we derive in the infrared limit a non-linear SU($N$) 
sigma model with a baby-Skyrme-type term on~$\Sigma^2$, which can be reduced further to $A_{N-1}$ Toda theory.
\end{abstract}

\end{center}
\end{titlepage}

\section {Introduction and summary}

\noindent 
The famous Alday-Gaiotto-Tachikawa (AGT) 2d-4d correspondence~\cite{1} relates
Liouville field theory on a punctured Riemann surface $\Sigma^2$ and SU(2) super-Yang--Mills (SYM)
theory  on a four-dimensional manifold $M^4$ (see e.g.~\cite{2} for a nice review and
references). This correspondence was quickly extended to 2d $A_{N{-}1}$ Toda field theory and
4d SU($N$) SYM~\cite{3}. Since then various AGT-like correspondences between theories on $n$-
and $(6{-}n)$-dimensional manifolds were investigated (see e.g.~\cite{4, 5}) for reviews and references).
One way to interpret these correspondences is to start from 6d $\cal N$=(2,0) supersymmetric conformal
field theory (CFT) on $M^n\times M^{6{-}n}$. The theory on $M^n$ would appear as a low-energy limit
when $M^{6{-}n}$ shrinks to a point, while the theory on $M^{6{-}n}$ would emerge when $M^n$ is scaled down 
to a point~\cite{2, 4, 5}. To be more precise, the AGT correspondence for $n{=}4$ relates the partition function
of 6d CFT of type $A_{N{-}1}$ on $M^4\times \Sigma^2$ with partition functions of theories on $M^4$
and $\Sigma^2$, which are all equal due to the conformal invariance of the 6d theory. 
The standard way of establishing such correspondences consists of calculating partition/correlation functions 
on $M^n$ and $M^{6{-}n}$ and to compare them. 
However, we are not aware of a {\it geometric\/} derivation of the AGT correspondence. 

Recently, a derivation of the AGT correspondence via reduction of 6d CFT to $M^4$ and $\Sigma^2$ was
discussed in~\cite{5a,6,7,8}. A relation between $A_{N{-}1}$ 6d CFT compactified on a circle and 5d SU($N$) SYM
was employed as well as further reduction to complex Chern-Simons theory on $\Sigma^2\times I$, where $I\subset M^4$
is an interval. Then a generalized version of the correspondence between 3d Chern-Simons theory and 2d CFT
was used, with the Nahm-pole boundary conditions translating to the constraints reducing WZW models to Toda
theories~\cite{6,7,8}.

In this paper we consider SU($N$) Yang--Mills theories on manifolds $M^4\times\Ical$ and
$M^3\times \Sigma^2$, where $M^4=M^3\times \tilde S^1$ is Lorentzian and $\Sigma^2\cong S^1\times\Ical$ is a two-sphere with two punctures.
The two manifolds agree in $M^3\times\Ical$ but differ in the additional circle $\tilde{S}^1$ versus $S^1$.
The supersymmetric extensions of Abelian gauge theories on these two 5d manifolds originate from 
type $A_{N-1}$ 6d $\Ncal = (2,0)$ supersymmetric CFT on $M^4\times\Sigma^2$, compactified on $S^1$ or $\tilde{S}^1$ 
with radii $R_1=\ve R_0$ or $\tR_1=\tilde\ve R_0$, respectively~\cite{Se, Sei, Dou}.
Choosing $\tilde\ve=\ve^{-1}$ the two 5d theories become dual to each other, and their gauge coupling constants $e^2$ or $\tilde{e}^2$ 
(in units of the reference scale~$R_0$) will be inversely proportional to one another, which is consistent with the arguments from~\cite{9}.
In order to have nontrivial vacuum solutions we take $M^3=\R_t\times T_p^2$ with a temporal direction~$\R_t$ and a one-punctured 2-torus~$T_p^2$.  

For further low-energy limits we introduce two scales, $\tR$ and $R$, which determine the sizes of the two factors in the product manifolds
$M^4_{\tR}\times\Ical_R$ and $M^3_{\tR}\times\Sigma^2_R$, respectively. In the first case, via $(\tR,R)=(R_0,\ve R_0)$ and $\ve\ll 1$ 
we shrink $\Ical_R$ to a point. Depending on the choice of reduction -- translational invariance along $\Ical_R$ or adiabatic 
approach~\cite{10,14,15,16,17,4} -- one obtains ($\Ncal\ge 2$ extended) Yang--Mills theory or the Skyrme model on the manifold $M^4_{R_0}$.
In the second case, taking $(\tR,R)=(\tilde\ve R_0,R_0)$ and $\tilde\ve\ll 1$ we scale down the metric on $\R_t$ and $T_p^2$.
The adiabatic method~\cite{14}-\cite{12} then produces a non-linear sigma model with a baby-Skyrme-type term on $\Sigma^2_{R_0}$.
Finally, we briefly discuss a reduction of this 2d SU($N$) sigma model to Toda field theory on $\Sigma^2_{R_0}$.

To summarize, we propose a geometric background for establishing 4d-2d AGT correspondences between field theories on $M^4$ and $\Sigma^2$.
We argue that these correspondences depend not only on the topology of $M^4$ and $\Sigma^2$ but also on the method employed for deriving 
the low-energy effective field theories in the infrared (using symmetries, adiabatic limit, constraints etc.).

\bigskip

\section {Two dual ways from 6d to 5d Yang--Mills theory}

\noindent {\bf 6d manifold.}
We consider the purported 6d Abelian $\Ncal = (2,0)$ CFT on the manifold
\begin{equation} \label{6metric}
M^4 \times \Sigma^2 \= M^3_\tR \times \tilde{S}^1_{\tR_1} \times S^1_{R_1} \times {\cal I}_R\ ,
\qquad\textrm{where}\quad M^3_\tR=\R_t\times T_p^2
\end{equation}
is a temporal cylinder over a two-torus with size $\tR$ and a puncture~$p$.
The radii of the circles $\tilde{S}^1$ and $S^1$ and the length of the interval ${\cal I}$ are indicated. 
The metric on this space is taken as
\begin{equation}
\begin{aligned}
\diff s^2_6 &\= \eta_{ab}\,\diff x^a \diff x^b + (\diff x^3)^2 + (\diff x^4)^2 + (\diff x^5)^2 \\[4pt]
&\= \tR^2\bigl[ -(\diff t)^2 + (\diff\a)^2+(\diff\b)^2\bigr] + \tR_1^2(\diff\vp)^2 + R_1^2(\diff\th)^2 + R^2(\diff z)^2\ .
\end{aligned}
\end{equation}
where $a,b=0,1,2$, and all spatial coordinates in the second line range over $[-\pi,\pi]$, with the Greek ones being periodic.
The coordinate relations are
\begin{equation}
x^0=\tR\,t\ ,\quad x^1=\tR\,\a\ ,\quad x^2=\tR\,\b\ ,\quad x^3=\tR_1\vp\ ,\quad x^4=R_1\th\ ,\quad x^5=R\,z\ .
\end{equation}

\medskip

\noindent {\bf Reduction to $M^4\times{\cal I}$.}
Let us equate $\tR_1=\tR$ and scale down $S^1$ by choosing
\begin{equation}
R_1 = \ve R_0 \ll R_0 \quad\with\quad \ve \ll 1
\end{equation}
relative to some fixed radius~$R_0$. 
This reduction gives rise to a 5d super-Maxwell theory~\cite{9} on 
\begin{equation}
M^4_\tR\times\Ical_R \= \R_t\times T_p^2 \times \tilde{S}^1_\tR \times {\cal I}_R
\end{equation}
with a metric
\begin{equation} \label{5Ametric}
\diff s^2_R \= \eta_{\ah\bh}\,\diff x^{\ah}\diff x^{\bh} + R^2(\diff z)^2
\quad\with\quad \ah,\bh\in\{0,1,2,3\}\ .
\end{equation}
It has a non-Abelian SU($N$) extension consistent with gauge invariance and supersymmetry, whose 
pure-gauge sector we shall consider further.
It is conjectured that this 5d SYM is the dimensional reduction of a non-Abelian 6d CFT.\footnote{
There are various mathematical approaches to such 6d non-Abelian CFTs, see e.g.~\cite{SW, SS} and references therein.}

\medskip

\noindent {\bf Gauge fields on $M^4\times{\cal I}$.}
The gauge potential  $\Acal$ (connection) and the gauge field $\Fcal$  (curvature)
both take values in the Lie algebra $su(N)$. On $M^4_\tR\times\Ical_R$ we have
\begin{equation}\label{2.2}
\Acal \= \Acal_{\ah}\,\diff x^{\ah}+\Acal_{z}\,\diff z\und
\Fcal \=\sfrac12\Fcal_{\ah\bh}\,\diff x^{\ah} \wedge \diff x^{\bh} + \Fcal_{\ah z}\,\diff x^{\ah} \wedge \diff z\ .
\end{equation}
For the generators $I_i$ in the fundamental $N\times N$ representation of SU($N$) we use the normalization
$\tr (I_iI_j) = -\sfrac12\de_{ij}$ for $i,j=1,\ldots, N^2{-}1$ and write $\Acal=\Acal^iI_i$ and $\Fcal=\Fcal^iI_i$.

For the metric tensor (\ref{5Ametric}) we have $g^{\ah\bh}_R=\eta^{\ah\bh}$ and $g^{zz}_R=R^{-2}$.
It follows that 
\begin{equation}
\Fcal^{\ah\bh}_R=\Fcal^{\ah\bh} \und \Fcal^{\ah z}_R=R^{-2}\Fcal^{\ah z}\ .
\end{equation}
The standard Yang--Mills action functional with this metric takes the form
\begin{equation}\label{2.3}
S_5 \= -\frac{R}{8e^2}\int_{M^4\times \Ical} \!\!\!\!\!\!\diff^4x\,\diff z\ 
\tr\bigl(\Fcal_{\ah\bh}\Fcal^{\ah\bh}+ \sfrac{2}{R^2}\Fcal_{\ah z}\Fcal^{\ah z}\bigr)\ ,
\end{equation}
where $e$ is the gauge coupling constant. 
It is known~\cite{9} that $e^2$ is proportional to the $S^1$ compactification radius $R_1{=}\ve R_0$.
A full supersymmetric extension of (\ref{2.3}) can be found e.g.~in \cite{4}.

\medskip

\noindent {\bf Reduction to $M^3\times\Sigma^2$.}
Alternatively, let us put $R_1=R$ and shrink $\tilde{S}^1$ by taking\begin{equation}
\tR_1 = \tilde\ve R_0 \ll R_0 \quad\with\quad \tilde\ve \ll 1\ .
\end{equation}
This reduction produces the 5d super-Maxwell theory on
\begin{equation}
M^3_\tR\times\Sigma^2_R \= \R_t\times T_p^2 \times S^1_R \times {\cal I}_R
\end{equation}
with a metric
\begin{equation} \label{5Bmetric}
\diff \tilde{s}^2_\tR \= \tR^2\bigl[ -(\diff t)^2 + (\diff\a)^2+(\diff\b)^2\bigr] 
+ \delta_{\bar a\bar b}\,\diff x^{\bar a}\diff x^{\bar b}
\quad\with\quad {\bar a},{\bar b}\in\{4,5\}\ .
\end{equation}

\medskip

\noindent {\bf Gauge fields on $M^3\times\Sigma^2$.}
As before, we extend the gauge group to SU($N$) and restrict ourselves to the pure Yang--Mills subsector since supersymmetry plays no role
in our discussion. Both gauge potential $\Acal$ and gauge field $\Fcal$ take values in $su(N)$, and on $M^3_\tR\times\Sigma^2_R$ one has
\begin{equation}\label{4.5}
\Acal \= \Acal_a\,\diff x^a+\Acal_{\bar a}\,\diff x^{\bar a}\und
\Fcal \=\sfrac12\Fcal_{ab}\,\diff x^a \wedge \diff x^b + \Fcal_{a\bar b}\,\diff x^a \wedge \diff x^{\bar b}
+\sfrac12 \Fcal_{\bar a \bar b}\,\diff x^{\bar a} \wedge \diff x^{\bar b}\ .
\end{equation}
Abbreviating the rescaled $M^3_\tR$ coordinates as $(t,\a,\b)=(y^\mu)$ with $\mu=0,1,2$, for the metric tensor (\ref{5Bmetric}) 
we have $\tilde{g}^{\mu\nu}_\tR=\tR^{-2}\eta^{\mu\nu}$ and $\tilde{g}^{\bar a \bar b}_\tR=\de^{\bar a \bar b}$.
Hence, for the upper components of $\Fcal$ we obtain
\begin{equation}
\Fcal^{\mu\nu}_\tR=\tR^{-4}\Fcal^{\mu\nu}\ ,\qquad
\Fcal^{\mu\bar b}_\tR=\tR^{-2}\Fcal^{\mu\bar b} \und
\Fcal^{\bar a \bar b}_\tR=\Fcal^{\bar a\bar b} \ .
\end{equation}
Then the standard Yang--Mills action functional on $M^3_\tR\times\Sigma^2_R$ will have the form
\begin{equation}\label{4.6}
\tilde{S}_5 \= -\frac{\tR^3}{8\tilde e^2}\int_{M^3\times \Sigma^2} \!\!\!\!\!\!\diff^3y\,\diff^2x\ \tr\bigl(
\sfrac{1}{\tR^4}\Fcal_{\mu\nu}\Fcal^{\mu\nu}+ \sfrac{2}{\tR^2}\Fcal_{\mu\bar b}\Fcal^{\mu\bar b}+\Fcal_{\bar a\bar b}\Fcal^{\bar a\bar b} \bigr)\ ,
\end{equation}
where $\tilde e$ is the dual gauge coupling, whose square is proportional to the $\tilde{S}^1$ compactification radius
$\tilde{R}_1=\tilde\ve R_0$. 
The reduction to $M^3_\tR\times\Sigma^2_R$ is dual to the one to $M^4_\tR\times\Ical_R$ in the sense that we may take $\tilde\ve=\ve^{-1}$
so that $\ve\ll1$ is related to $\tilde\ve\gg1$ and vice versa. Hence the Yang--Mills theories (\ref{2.3}) and (\ref{4.6}) are dual in
agreement with the discussion in~\cite{9}.

\bigskip

\section{Extended SYM and Skyrme model on $M^4$ from SYM on $M^4\times\Ical$}

\noindent {\bf SYM on $M^4$.} 
In the SYM model on $M^4_\tR\times\Ical_R$, discussed in the previous section as the first reduction from 6d,
there exist two types of further reduction to four dimensions, both of which appear in the literature as
low-energy limits: reduction with respect to the translation $\pa_z$~\cite{9} and reduction via the adiabatic approach
(see e.g.~\cite{10, 4, 11, 12}). Both reductions shrink $\Ical_R$ to a point, via
\begin{equation} \label{infraredA}
R=\ve R_0\ll R_0 \und \tR=R_0 \quad\for \ve\ll 1\ ,
\end{equation}
but the final result is different. We drop the subscript on $M^4_{\tR}$ and treat both cases in turn.

Let us impose translation invariance along $\Ical_R$ on all fields,
\begin{equation}\label{3.1}
\pa_z\Acal_a =0 = \pa_z \Phi \quad\for \Phi :=\Acal_5 = \sfrac1R\Acal_{z}\ ,
\end{equation}
by saying that for $R\ll\tR$ the momenta along $\Ical_R$ are much larger than along $M^4$. 
Then one can discard all higher modes of $\Acal_a$ and $\Phi$ as well as of scalar
and spinor fields of maximally supersymmetric Yang--Mills theory on $M^4\times\Ical_R$.

After substituting (\ref{3.1}), the Yang--Mills action (\ref{2.3}) reduces to
\begin{equation}\label{3.2}
S_4\=-\frac{\pi R}{4e^2}\int_{M^4} \!\!\diff^4x\ \tr\bigl(\Fcal_{ab}\Fcal^{ab} + 2D_{a}\Phi D^{a}\Phi\bigr)\ ,
\end{equation}
where $D_a =\pa_a + [\Acal_a, \ \cdot\ ]$. Likewise, the full SYM theory on $M^4\times\Ical_R$
passes to $\Ncal =2$ or  $\Ncal =4$ super-Yang--Mills theory on $M^4$, depending on twisting along $\Sigma^2$ 
and other assumptions (see e.g.~\cite{2,9}). The discussion simplifies because in the 6d to 5d reduction
the two-punctured two-sphere $\Sigma^2\cong S^1_{R_1}\times\Ical_R$ gets deformed into a thin cylinder as in \cite{4, 10}.

\medskip

\noindent {\bf Skyrme model.} 
A different infrared limit of SYM theory on $M^4\times\Ical_R$, discussed e.g.~in~\cite{4,10,11}, introduces
\begin{equation}\label{3.3}
h(z) \= \Pcal\exp\biggl(\int_{{-}\pi}^z \Acal_y\diff y\biggr )\und g=h(z{=}\pi )\in {\mathrm {SU}}(N)\ ,
\end{equation}
where $\Pcal$ denotes path ordering. The group element $g$ is the holonomy of $\Acal$ along $\Ical$.
In the low-energy limit, when the length of $\Ical$ becomes small, the 5d YM
theory~(\ref{2.3}) reduces to the Skyrme model on $M^4$,
\begin{equation}\label{3.4}
S_{\textrm{eff}} \= -\int_{M^4}\!\!\diff^4x\ \biggl\{\frac{f^2_\pi}{4}\,\eta^{ab}\, \tr (L_a L_b)\ +\
\frac{1}{32\,\vs^2}\,\eta^{ac}\eta^{bd}\,\tr\bigl([L_a, L_b][L_c, L_d]\bigr)\biggr\}\ ,
\end{equation}
where $L_a=g^{-1}\pa_a g$ for $g$ from (\ref{3.3}), $\vs$ is the dimensionless Skyrme parameter,
and $f_{\pi}$ is interpreted as the pion decay constant.
Their relation to the dimensionful 5d gauge coupling $e^2$ and the infrared scale~$R$ is
\begin{equation}\label{3.5}
\frac{f_\pi^2}{4} = \frac{\pi}{4\,e^2 R}  \und  \frac{1}{32\,\vs^2} = \frac{\pi R}{120\,e^2}\ .
\end{equation}
Recall that $e^2$ is proportional to $R$~\cite{9} and $R=\ve R_0$ becomes small,
so that $f_\pi^2\sim\ve^{-2}$ and $\vs^{-2}\sim\ve^0$, and higher-order (in~$R$) terms are suppressed.

The derivation of (\ref{3.4}) does not impose (\ref{3.1}). It is based on the adiabatic
approach~\cite{14}-\cite{17} which is equivalent to the moduli-space approximation with moduli given by
$g=g(x)$ from (\ref{3.3}). The two-derivative term in (\ref{3.4}) is the standard 4d non-linear sigma model,
and its supersymmetric version was derived from 5d in~\cite{4}. The four-derivative term in (\ref{3.4})
stabilizes solitons against scaling. This term was deduced from 5d SYM on $M^4\times\Ical_R$ in~\cite{11}, 
where its possible supersymmetrization, yet unknown, was briefly discussed.

Thus, in the infrared of 5d SYM theory on $M^4\times\Ical_R$ one can find two different models on $M^4$:
$\Ncal\ge 2$ SYM theory and the Skyrme model. The latter describes low-energy QCD by interpreting mesons as
fundamental and baryons as solitons. Away from the infrared limit the Skyrme model gets extended by
an infinite tower of heavy mesons beyond the leading Skyrme term displayed in~(\ref{3.4})~(cf.~\cite{SaSu, Sut}).

\bigskip

\section{Moduli space of YM vacua on $M^3$}

\noindent 
For the remainder of the paper we focus on the second kind of 6d to 5d reduction presented in Section~2,
namely the dual action~(\ref{4.6}), and its own infrared simplification from $M^3_\tR\times\Sigma^2_R$ to~$\Sigma^2$.
In a manner dual to~(\ref{infraredA}), we put
\begin{equation}
\tR = \tilde\ve R_0 \ll R_0 \und R = R_0 \quad\for \tilde\ve\ll 1\ ,
\end{equation}
which scales down the metric~(\ref{5Bmetric}) in the $M^3_\tR$ direction.
We drop the reference to the (now fixed) scale of~$\Sigma^2_R$.
Hence, for $\tR\ll R$ the 5d YM theory reduces to an effective 2d field theory on $\Sigma^2$ which we now describe.

\medskip

\noindent {\bf  Flat connections on $M^3$.} 
Our derivation of the low-energy limit employs the adiabatic method~\cite{14}-\cite{19}
(for brief reviews and more references see e.g.~\cite{22, 20}). 
In this approach one firstly restricts to $M^3_\tR$ by putting 
\begin{equation}\label{5.0}
\Acal_{\bar a}=0 \quad\for \bar a\in\{4,5\}
\end{equation}
and classifies the classical solutions $\Acal_m(y^n)$ on $M^3_\tR$, independent of the $\Sigma^2$ coordinates~$x^{\bar a}$.
Secondly, one declares that the moduli~$X^\a$, which parametrize such solutions, become functions of $x^{\bar a}\in\Sigma^2$. 
Thirdly, one introduces small fluctuations $\delta\Acal_m$ and $\Acal_{\bar a}$ depending on $y^n\in M^3_\tR$ and on the moduli 
functions $X^\a (x^{\bar a})$, substitutes them into (\ref{4.6}) and obtains an effective field theory for $X^\a$ on~$\Sigma^2$. 
Since in the following we want to study small fluctuations around the vacuum manifold, 
we first take a look at the vacuum configurations.

The vacua on $M^3_\tR$ are given by
\begin{equation}\label{5.1}
\Fcal_{\mu\nu} =0 \ ,
\end{equation}
meaning flat connections on~$M^3_\tR$. The consideration of flat connections~(\ref{5.1}) guarantees that the action~(\ref{4.6}) 
will be nonsingular in the limit $\tilde\ve\to0$ ($\tR\to0$). One can always choose the gauge $\Acal_0 =0$, which simplifies~(\ref{5.1}) to
\begin{equation}\label{5.2}
\Fcal_{mn}=0 \quad\Leftrightarrow\quad \Fcal_{T^2_p}=0 \quad\for m,n\in\{1,2\} \ ,
\end{equation}
describing flat connections $\Acal_{T^2_p}$ on the punctured torus $T^2_p=T^2{\setminus}\{p\}$.

\medskip

\noindent {\bf Flat connection on $T^2_p$.} 
It is well known that gauge bundles over smooth tori $T^2$ (compact, without punctures) admit only reducible flat connections~\cite{23,17}.
However, this theorem is not valid on Riemann surfaces with punctures or fixed points (see e.g.~\cite{24,25,26}).
In particular, the moduli space $\Mcal_{T^2_p}$ of flat connections on a $K$-bundle over $T^2_p$ is the gauge group~$K$,
\begin{equation}\label{5.3}
\Mcal_{T^2_p}\=\Ncal_{T^2_p}/\Gcal_{T^2_p}\=\sfrac{\{\textrm{flat connections}\}}{\{\textrm{gauge transformations}\}} \= K\ ,
\end{equation}
where the group of gauge transformations $\Gcal_{T^2_p}=C^\infty(T^2_p, K)$ forms the fibres over the points in $\Mcal_{T^2_p}$ for the bundle
\begin{equation}\label{5.4}
\pi :\ \Ncal_{T^2_p}\ \stackrel{\Gcal_{T^2_p}}{\longrightarrow}\  \Mcal_{T^2_p}=K\ .
\end{equation}
We specialize to $K=\textrm{SU}(N)$.

\medskip

\noindent {\bf Flat variation of $\Acal^{}_{T^2_p}$}. 
Any solution  $\Acal^{}_{T^2_p}=\Acal_m\diff y^m$ of (\ref{5.2}) is parametrized by the coordinates $X^\a$ on $\Mcal^{}_{T^2_p}=\textrm{SU}(N)$, i.e.
\begin{equation}
\Acal_m = \Acal_m(y^n,X^\a)\ .
\end{equation}
In general, flat connections  $\Acal^{}_{T^2_p}$ belong to the space $\Ncal^{}_{T^2_p}$ fibred over $\Mcal^{}_{T^2_p}$, as we have the bundle (\ref{5.4}). 
Hence, flat variations live in the tangent bundle $T\Ncal^{}_{T^2_p}$, defined as the fibration
\begin{equation}\label{5.15}
\pi_* :\quad T\Ncal^{}_{T^2_p}\ \longrightarrow\ T\Mcal^{}_{T^2_p}
\end{equation}
with fibres $T^{}_{\Acal_{T^2_p}}\Gcal_{T^2_p}\cong\,$Lie$\Gcal^{}_{T^2_p}$ at any point  $\Acal^{}_{T^2_p}\in \Ncal^{}_{T^2_p}$. We have
\begin{equation}\label{5.16}
T^{}_{\Acal_{T^2_p}}\Ncal^{}_{T^2_p}\=\pi^*T^{}_{\Acal^{}_{T^2_p}} \Mcal_{T^2_p}\oplus T^{}_{\Acal_{T^2_p}}\Gcal^{}_{{T^2_p}}
\ \cong\ \mfrak_+\oplus {\rm Lie}\,\Gcal^{}_{{T^2_p}}\ .
\end{equation}
The variation of $\Acal_m$ along $T\Mcal^{}_{T^2_p}$ is then given by
\begin{equation}\label{5.17} 
\de_\a \Acal_m \= \pa_\a\Acal_m - D_m\eps_\a\ ,
\end{equation}
where $\eps_\a$ is a suitable gauge parameter which brings $\pa_\a\Acal_m$ back to $\pi^*T^{}_{\Acal^{}_{T^2_p}}\Mcal_{T^2_p}$.
This makes sure that the variation $\delta_\a\Acal_m$ obeys a linearization of the flatness condition~(\ref{5.1}),
\begin{equation} \label{linflat}
D_m \delta_\a\Acal_n - D_n \delta_\a\Acal_m \= 0\ .
\end{equation}

\medskip    

\noindent {\bf SU($N$) as coset space.} 
We would like to realize SU($N$) as a coset $G/H$. The simplest way is 
\begin{equation}\label{5.5}
G\=\mbox{SU}_+(N)\times \mbox{SU}_-(N)\und H\=\mbox{diag (SU}_+(N)\times \mbox{SU}_-(N))\ .
\end{equation}
Accordingly, $\gfrak =\hfrak\oplus \mfrak$, where $\hfrak =\,$Lie$H$. At any point in $G/H$ the tangent space is isomorphic to $\mfrak$.
Let $\{I_i\}$ be a basis of the Lie algebra $\gfrak$ realized as $2N\times 2N$ block-diagonal matrices with normalization 
$\tr (I_iI_j)=-\sfrac12\de_{ij}$ for $i,j=1,\ldots,2(N^2{-}1)$. 
We choose bases $\{I_{\hat{\imath}}\}$ for $\hfrak$ and $\{I_{\bar{\imath}}\}$ for $\mfrak$ so that $\{I_i\}=\{I_{\hat{\imath}},I_{\bar{\imath}}\}$, 
and both $\hat{\imath}$ and $\bar{\imath}$ range from 1 to $N^2{-}1$.
Note that in general $G/H$ is not symmetric, and we consider this case of group manifolds with torsion.

The coset $G/H=\textrm{SU}(N)$ supports an orthonormal frame $\{e^{\bar{\imath}}\}$ of one-forms locally providing the $G$-invariant metric
\begin{equation}\label{5.6}
\diff s^2_{{\rm{SU}}(N)} \= \de_{\bar{\imath}\bar{\jmath}}\,e^{\bar{\imath}}e^{\bar{\jmath}}
\=\de_{\bar{\imath}\bar{\jmath}}\,e^{\bar{\imath}}_\a e^{\bar{\jmath}}_\b\,\diff X^\a \diff X^\b
\ =:\ g_{\a \b}\,\diff X^{\a} \diff X^{\b} \quad\for \a,\b =1,\ldots,N^2-1\ ,
\end{equation}
where $\{X^\a\}$ is a set of local coordinates of a point $X\in G/H=\textrm{SU}(N)$ 
and $\pa_\a=\frac{\pa}{\pa X^\a}$ will denote derivatives with respect to them.

There is the natural projection
\begin{equation}\label{5.7}
q:\quad G\stackrel{H}{\longrightarrow} G/H\ ,
\end{equation}
and on the principal $H$-bundle (\ref{5.7}) over $G/H=\textrm{SU}(N)$ there exists a one-parameter family of connections
\begin{equation}\label{5.8}
\Acal^\vk_{{\rm{SU}}(N)}\=\vk\, e^{\hat\imath}I_{\hat\imath}\=\vk\, e^{\hat\imath}_\a I_{\hat\imath}\,
\diff X^\a\ =:\ \Acal_\a\,\diff X^\a \quad\for \vk\in\R
\end{equation}
with curvature
\begin{equation}\label{5.9}
\Fcal^\vk_{{\rm{SU}}(N)}\=\sfrac12\,\vk(\vk{-}1)\,f^{\hat\imath}_ {\hat{\jmath}\hat k}\, e^{\hat\jmath}\wedge e^{\hat k} 
\= \sfrac12\,\Fcal_{\a\b}\,\diff X^\a\wedge\diff X^\b\ .
\end{equation}
The one-forms $e^i=(e^{\hat\imath}, e^{\bar\imath})$ on $G/H$ are pulled back from $G$.
We choose 
\begin{equation}
\mfrak \= \mfrak_+\ \equiv\ (m,0)\ \subset\ su_+(N)\oplus su_-(N)\ ,
\end{equation}
so that the Maurer-Cartan equations have the form
\begin{equation}\label{5.10}
\diff e^{\bar\imath} \= -f^{\bar\imath}_{\hat{\jmath}\bar k}\,e^{\hat\jmath}\wedge e^{\bar k}
-\sfrac12\,f^{\bar\imath}_ {\bar{\jmath}\bar k}\,e^{\bar\jmath}\wedge e^{\bar k}
\und 
\diff e^{\hat\imath} \= -\sfrac12\,f^{\hat\imath}_{\hat{\jmath}\hat k}\,e^{\hat\jmath}\wedge e^{\hat k}\ .
\end{equation}
The second relation in (\ref{5.10}) was used to derive~(\ref{5.9}). For more details and references see~\cite{27}.

\bigskip

\section{Toda theory on $\Sigma^2$ from YM on $M^3\times \Sigma^2$}

\noindent{\bf Moduli space approximation.} 
In the adiabatic approach, applicable for $\tilde\ve\ll 1$, the moduli approximation assumes that the moduli $X^\a$ 
vary with the coordinates $x^{\bar a}\in \Sigma^2$~\cite{14}-\cite{19}. In this way, the moduli of flat connections on $T^2_p$ define a map
\begin{equation}\label{6.1}
X\ :\quad \Sigma^2\ \to\ \mbox{SU}(N)\ ,
\end{equation}
so that $X^\a = X^\a(x^{\bar a})$ may be considered as dynamical fields. 
Admitting $x^{\bar a}$ dependence only via $X^\a\in\textrm{SU}(N)$, our fields take the form
\begin{equation}\label{6.2}
\Acal_m \= \Acal_m\bigl(y^n,X^\a(x^{\bar b})\bigr) \und \Acal_{\bar a} \= \Acal_{\bar a}\bigl(y^n,X^\a(x^{\bar b})\bigr)\ ,
\end{equation}
where we stay with the gauge choice $\Acal_0=0$.

We are interested in the low-energy effective action for~$X^\a$ on~$\Sigma^2$, generated by small fluctuations~$(\delta\Acal_m,\Acal_{\bar a})$
around the vacuum (\ref{5.0}) and~(\ref{5.1}) which respect the flatness on $T_p^2$, i.e.~obey~(\ref{linflat}). 
This condition keeps the infrared-singular first term in~(\ref{4.6}) removed.
We need to compute the field strength components $\Fcal_{\bar a m}$ and $\Fcal_{\bar a\bar b}$ of the vacuum deviations (\ref{6.2}). 
The mixed ones yield
\begin{equation}
\begin{aligned}
\Fcal_{\bar a m} \= \pa_{\bar a}\Acal_m-D_m\Acal_{\bar a}
&\= (\pa_{\bar a} X^\a)\bigl(\pa_\a\Acal_m-D_m\Acal_\a\bigr) \\[4pt]
&\= (\pa_{\bar a} X^\a)\bigl(\delta_\a\Acal_m+D_m(\eps_\a-\Acal_\a)\bigr)
\= (\pa_{\bar a} X^\a)\,\delta_\a\Acal_m\ ,
\end{aligned}
\end{equation}
where (\ref{5.17}) was employed, 
and we put $\eps_\a=\Acal_\a$ to keep $\Fcal_{\bar a m}$ tangent to the vacuum moduli space~$\Mcal_{T^2_p}$. 
The moduli-space connection $\Acal_\a\diff X^\a=\Acal^\vk_{{\rm{SU}}(N)}$ has been computed in~(\ref{5.8}).
The fact that we choose $\eps_{\alpha}\diff X^\a$ equal to  $\Acal^\vk_{{\rm{SU}}(N)}$ which is constant on~$T_p^2$, 
i.e.~it does not depend on $x^m$, does not matter because in the infrared limit $\tR\ll 1$ only the zero mode 
in a Fourier expansion of the $x^m$ dependence on~$T_p^2$ will survive.
As a result, we arrive at the infrared approximation
\begin{equation} \label{6.11}
\begin{aligned}
\Acal_{\bar a}&\=(\pa_{\bar a} X^{\a})\,\Acal_{\a} \=
\vk\,(\pa_{\bar a} X^{\a})\,e_{\a}^{\hat\imath}I_{\hat\imath}\=\vk\,g^{-1}\pa_{\bar a}g \= \vk\,L_{\bar a}\ ,\\[4pt]
\Fcal_{\bar a\bar b} &\= \vk(\vk{-}1)\,(\pa_{\bar a}X^\a)(\pa_{\bar b} X^\b)\,f^{\hat\imath}_{\hat\jmath\hat k}e^{\hat\jmath}_\a e^{\hat k}_\b\,I_{\hat\imath}
\= \vk(\vk{-}1) [L_{\bar a}, L_{\bar b}]\ ,
\end{aligned}
\end{equation}
where $g\in\textrm{SU}(N)$ and we abbreviated
\begin{equation}
L_{\bar a} \= (\pa_{\bar a} X^\a)\,g^{-1}\pa_\a g \= g^{-1}\pa_{\bar a} g\ .
\end{equation}

\medskip

\noindent {\bf Non-linear sigma model.} 
Substituting $\Fcal_{\bar a m} =  (\pa_{\bar a} X^\a)\,\de_{\a} \Acal_m$ into the action (\ref{4.6}), the second term becomes
\goodbreak
\begin{equation}\label{6.6}
\begin{aligned}
S_{\textrm{kin}} &\=
-\frac{\tR}{4\,\tilde e^2}\int_0^T \!\!\!\diff t \int_{T^2_p} \!\!\!\diff^2y \int_{\Sigma^2} \!\!\diff^2x\,
\tr(\Fcal_{\bar a m}\Fcal^{\bar a m}) \\
&\= \frac{\tR\,T}{2\,\tilde e^2}\int_{\Sigma^2} \!\!\!\diff^2x\,\de^{\bar a\bar b}\,\pa_{\bar a} X^\a\pa_{\bar b} X^\b\,g_{\a\b}
\=-\frac{\tR\,T}{4\,\tilde e^2}\int_{\Sigma^2}\!\!\!\diff^2x\,\de^{\bar a\bar b}\,\tr (L_{\bar a} L_{\bar b})\ ,
\end{aligned}
\end{equation}
where we have compactified the time direction with a finite length~$T\tR=\tilde\ve T R_0$ and
\begin{equation}\label{6.8}
g_{\a\b} \= -\sfrac12\,\int_{T^2_p}\!\!\!\diff^2y\,\de^{mn}\,\tr(\de_{\a} \Acal_{m}\de_{\b} \Acal_{n})
\=\de_{\bar\imath\bar\jmath}\,e_\a^{\bar\imath}e_\b^{\bar\jmath}
\=-\sfrac12\,\tr\bigl[(g^{-1}\pa_\a g)(g^{-1}\pa_\b g)\bigr]
\end{equation}
is the moduli-space metric (\ref{5.6}) on the group SU($N$).
Thus, this part of the action  (\ref{4.6}) reduces to the standard non-linear sigma model on $\Sigma^2$ with SU($N$) as target space.

\medskip

\noindent {\bf Baby-Skyrme-type term.} 
It remains to evaluate the last term in the action (\ref{4.6}). Substituting (\ref{6.11}) we obtain
\begin{equation}\label{6.12}
S_{\textrm{sky}}
\=-\frac{\tR^3}{8\,\tilde e^2}\int_0^T \!\!\diff t \int_{T^2_p} \!\!\!\diff^2y \int_{\Sigma^2} \!\!\!\diff^2x\,
\tr\bigl(\Fcal_{\bar a\bar b}\Fcal^{\bar a\bar b}\bigr)
\= -\frac{\pi^2 \tR^3 T}{\tilde e^2}\,\vk^2(\vk{-}1)^2\int_{\Sigma^2} \!\!\!\diff^2x\ \tr\bigl([L_{\bar 1}, L_{\bar 2}]^2\bigr)\ ,
\end{equation}
where we used the fact that area of $T^2_p$ is $4\pi^2\tR^2$. Hence, in the infrared $\tilde\ve\ll 1$ the Yang--Mills action (\ref{4.6}) reduces to the action
\begin{equation}\label{6.13}
\tilde{S}_{\textrm{eff}}\=S_{\textrm{kin}}+S_{\textrm{sky}}
\= -\int_{\Sigma^2}\!\!\diff^2x\,\biggl\{ \frac{\tilde{f}^2_\pi}{4}\,\delta^{\bar a\bar b}\,\tr (L_{\bar a} L_{\bar b})
+ \frac{1}{16\,\tilde\vs^2}\,\tr\bigl([L_{\bar 1}, L_{\bar 2}]^2\bigr)\biggr\}\ ,
\end{equation}
\begin{equation}
\textrm{with the assigments} \qquad
\frac{\tilde{f}_\pi^2}{4} = \frac{\tR\,T}{4\,\tilde{e}^2}  \und  \frac{1}{16\,\tilde\vs^2} = \frac{\pi^2 \tR^3 T}{\tilde{e}^2}\vk^2(\vk{-}1)^2\ .
\end{equation}
Recall that $\tilde{e}^2$ is proportional to $\tR$ and $\tR=\tilde\ve R_0$ becomes small, so that 
so that (for finite $T$) $\tilde{f}_\pi^2\sim\tilde\ve^0$ and $\tilde\vs^{-2}\sim\tilde\ve^2$, and higher-order (in~$\tR$) terms are suppressed.\footnote{
To keep a finite time interval $T\tR$ in the infrared, one should scale $T\sim\tilde\ve^{-1}$, which modifies this behavior.}
The term  (\ref{6.12}) looks unusual for 2d sigma models. However, if we take $N{=}2$ and restrict $g\in\,$SU(2) to
the coset $\C P^1=\,$SU(2)/U(1)$\,\subset\,$SU(2),
then exactly this term appears in the baby Skyrme model in two dimensions~\cite{28}.

\medskip

\noindent {\bf Toda model.} 
By choosing $\vk=1$ in (\ref{6.12}) one obtains the standard principal chiral model on $\Sigma^2\cong S^1\times\Ical$, 
i.e.~$\tilde{S}_{\textrm{eff}}=S_{\textrm{kin}}$. In~\cite{6} 6d $\Ncal{=}(2,0)$ CFT was reduced to Chern-Simons theory on $\Sigma^2\times I$
and then to a sigma model with torsion (WZW model) on $\Sigma^2$. Finally, by imposing current constraints, one can reduce the WZW model to Toda field
theory~\cite{29, 6, 7, 8}. Relations between 4d SYM theories and 2d WZW theories were considered earlier in~\cite{30, 31}.

The standard principal chiral model (\ref{6.6}) can also be reduced to Toda field theory on $\Sigma^2$. 
The 2d sigma model equations are {\it equivalent\/} to the two-dimensional self-dual Chern-Simons (CS) equations 
which by imposing some constraints can be {\it reduced\/} to the affine Toda equations~\cite{32, 33}.
Thus, we have the correspondence
\vspace{-3mm}
\begin{equation}\label{6.14}
\mbox{ 2d sigma model}\ \Leftrightarrow\ \mbox{ 2d self-dual CS}\ \longrightarrow\ \mbox{ 2d Toda theory}\ .
\end{equation}
This correspondence allows one, in particular, to construct solutions of the Toda field equations 
from uniton solutions of the SU($N$) sigma model as was discussed in~\cite{33}.


\bigskip

\noindent {\bf Acknowledgements}

\noindent
This work was partially supported by the Deutsche Forschungsgemeinschaft grant LE 838/13.
It is based upon work from COST Action MP1405 QSPACE, supported by COST (European Cooperation
in Science and Technology).


\end{document}